\documentclass[sigconf]{acmart}

\usepackage{booktabs}

\usepackage{pifont}
\newcommand{\cmark}{\ding{51}}

\usepackage{tikz}
\newcommand\copyrightnotice[1]{
	\begin{tikzpicture}[remember picture,overlay]
		\node[anchor=south,yshift=20pt] at (current page.south) {\fbox{\parbox{\dimexpr\textwidth-\fboxsep-\fboxrule\relax}{#1}}};
	\end{tikzpicture}
}

\AtBeginDocument{%
  \providecommand\BibTeX{{%
    \normalfont B\kern-0.5em{\scshape i\kern-0.25em b}\kern-0.8em\TeX}}}

\copyrightyear{2021}
\acmYear{2021}
\setcopyright{acmlicensed}\acmConference[ARES 2021]{The 16th International Conference on Availability, Reliability and Security}{August 17--20, 2021}{Vienna, Austria}
\acmBooktitle{The 16th International Conference on Availability, Reliability and Security (ARES 2021), August 17--20, 2021, Vienna, Austria}
\acmPrice{15.00}
\acmDOI{10.1145/3465481.3465749}
\acmISBN{978-1-4503-9051-4/21/08}

\begin{document}

\title{First Step Towards EXPLAINable DGA Multiclass Classification}

\author{Arthur Drichel}
\email{drichel@itsec.rwth-aachen.de}
\affiliation{%
	\institution{RWTH Aachen University}
	\city{}
	\country{}
%	\city{Aachen}
%	\country{Germany}
}
\author{Nils Faerber}
\email{nils.faerber@rwth-aachen.de}
\affiliation{%
	\institution{RWTH Aachen University}
	\city{}
	\country{}
}
\author{Ulrike Meyer}
\email{meyer@itsec.rwth-aachen.de}
\affiliation{%
	\institution{RWTH Aachen University}
	\city{}
	\country{}
}

\renewcommand{\shortauthors}{Drichel et al.}

% !TEX root = ../paper.tex
\begin{abstract}

Numerous malware families rely on domain generation algorithms (DGAs) to establish a connection to their command and control (C2) server. Counteracting DGAs, several machine learning classifiers have been proposed enabling the identification of the DGA that generated a specific domain name and thus triggering  targeted remediation measures. However, the proposed state-of-the-art classifiers are based on deep learning models. The black box nature of these makes it difficult to evaluate their reasoning. The resulting lack of confidence makes the utilization of such models impracticable. In this paper, we propose EXPLAIN, a feature-based and contextless DGA multiclass classifier. We comparatively evaluate several combinations of feature sets and hyperparameters for our approach against several state-of-the-art classifiers in a unified setting on the same real-world data. Our classifier achieves competitive results, is real-time capable, and its predictions are easier to trace back to features than the predictions made by the DGA multiclass classifiers proposed in related work.

\end{abstract}





\maketitle
% !TEX root = ../paper.tex
\section{Introduction}
\label{sec:introduction}

Modern botnets rely on domain generation algorithms (DGAs)  in order to enable the communication between the botnet herder and malware infected devices. In contrast to the usage of single fixed IP-addresses or fixed domain names, the communication of DGA-based malware is harder to block as they generate a vast amount of algorithmically generated domains (AGDs). The botnet herder is aware of the generation scheme and thus able to register a small subset of the generated domains in advance. The bots query all AGDs, one-by-one, trying to obtain the valid IP-address of their command and control (C2) server. Most of these queries result in non-existent domain (NXD) responses since the major fraction of the AGDs is not registered. Through an analysis of occurring NXDs within a network it is possible to detect DGA activities and thereby to take appropriate countermeasures even before the bots are commanded to partake in any malicious action. However, the detection of malicious AGDs is not trivial as NXDs can also be the product of typing errors, misconfigured or outdated software, or the intentional misuse of the DNS e.g. by antivirus software.  We refer to the task of  separating benign from malicious domain names as the DGA binary classification task. Going one step further, it is desirable to not only detect malicious AGDs but to attribute them to the specific DGAs that generated the domain names. This multiclass classification task narrows down the malware family and ultimately enables the execution of targeted remediation measures. The DGA multiclass classification task is far more complex as the classifier has to cope with an increased number of classes compared to the two classes in binary classification. For instance, the open-source threat intelligence feed of DGArchive~\cite{plohmann_comprehensive_2016} contains approximately 115 million unique ADGs generated by 92 different DGAs.

\copyrightnotice{\copyright\space Copyright held by the owner/author(s) 2021. This is the author's version of the work. It is posted here for your personal use. Not for redistribution. The definitive version was published in Proceedings of the 16th International Conference on Availability, Reliability and Security (ARES 2021), https://doi.org/10.1145/3465481.3465749}
In the past, a multitude of different approaches have been proposed in order to detect DGAs which can  broadly be grouped into context-aware and contextless approaches. The context-aware approaches (e.g. \cite{antonakakis_throwaway_2012,bilge_exposure_2014,grill_detecting_2015,yadav_winning_2012,schiavoni_phoenix_2014}) make use of additional contextual information next to the domain name that is to be classified trying to enhance the classification performance.  On the other side, contextless approaches (e.g. \cite{schuppen_fanci_2018,drichel_analyzing_2020,woodbridge_predicting_2016,yu_character_2018,saxe_expose_2017}) entirely rely on information extracted from a single domain name for classification and are thus less resource intensive and less privacy invasive. Prior studies (e.g.~\cite{schuppen_fanci_2018,drichel_analyzing_2020,woodbridge_predicting_2016,yu_character_2018})  have shown that the contextless approaches are able to achieve state-of-the-art performance while not depending on the extensive tracking of DNS traffic.

The proposed contextless machine learning classifiers can further be separated into feature-based, such as support vector machines (SVMs) or random forests (RFs) (e.g. \cite{schuppen_fanci_2018}), and deep learning based approaches, such as recurrent (RNNs), convolutional (CNNs), or residual neural networks (ResNets) (e.g.~\cite{drichel_analyzing_2020,woodbridge_predicting_2016,yu_character_2018,saxe_expose_2017}). For the multiclass task, however, only deep learning based contextless classifiers are available to date. While these classifiers achieve a very promising performance, deep learning classifier are often said to lead to less well explainable predictions compared to feature-based classifiers.

We intend this work to be a first step towards explainable DGA multiclass classification. Ultimately, this requires a focused comparison of different approaches to DGA multiclass classification with respect to explainability. To this end, competitively performing feature-based and deep learning based approaches have to be contrasted. However, to the best of our knowledge, there currently is no feature-based multiclass classifier and the ones that can easily be constructed from existing binary classifiers (see Section ~\ref{sec:selected_sota_classifiers}) do not exhibit competitive performance (see Section ~\ref{sec:evaluation1}). We thus, first focus on developing a feature-based multiclass classifier that achieves a classification performance comparable to previously proposed deep learning classifiers. We describe the engineering of our classifier in detail for the sake of transparency, which is a major requirement for explainability. Without a transparent feature engineering and selection process a feature-based classifier operates similar to a black box.

In this paper, we make the following contributions:

First, we propose EXPLAIN and publish the source code along with this paper.\footnote{https://gitlab.com/rwth-itsec/explain} To the best of our knowledge, this is the first contextless and feature-based machine learning classifier for DGA multiclass classification. The main target user groups for our classifier are security operation center (SOC) analysts and model developers. SOC analysts potentially benefit from the use of EXPLAIN over deep learning approaches when analyzing predictions in search for potential false positives or false negatives. Here, EXPLAIN enables analyzing the features that contributed to the classification result. Model developers, on the other hand, heavily profit from the adaptability of our approach. For instance, features can be adjusted or new discriminating features can be engineered in order to enable the correct attribution of newly discovered DGAs.

In addition to presenting EXPLAIN itself, we also present the methodology we used for engineering
this classifier. Thereby, we provide a methodical description which covers the complex process of feature engineering, feature selection,
and hyperparameter optimization. This methodology can be used
to create feature-based classifiers using domain knowledge for
different areas of application. Note that related work on feature-based machine learning
classifiers for DGA detection such as \cite{schuppen_fanci_2018,antonakakis_throwaway_2012} fall short in providing
such methodology. There, only the utilized features are presented without stating whether and how
feature selection and hyperparameter optimization were performed.
This not only renders the decisions made by the authors incomprehensible
but also hinders the reuse of the methodology which
might be of help also for different areas of application.

Second, we extensively compare our proposed classifier with
state-of-the-art deep learning based approaches. Here, we demonstrate
that our classifier is able to achieve results competitive to the currently best performing
deep learning classifiers.

% !TEX root = ../paper.tex
\section{Related Work}
\label{sec:related_work}

In the following, we first present related work which focuses on feature-based machine learning approaches for DGA detection as well as on contextless deep learning based classifiers. Afterwards we discuss which of the state-of-the-art classifiers we selected for our comparative evaluation and explain any modification we applied to the classifiers to enable a fair comparison.

\subsection{DGA Detection Classifiers}

\paragraph{\textbf{FANCI}}
Sch\"uppen et al.~\cite{schuppen_fanci_2018} proposed a system called  Feature-based Automated NXDomain Classification and Intelligence (FANCI) which is capable of separating benign from malicious domain names. For solving the DGA binary classification task FANCI implements an SVM and an RF-based classifier and makes use of 12 structural, 7 linguistic, and 22 statistical features.\footnote{FANCI uses 21 features, but feature \#20 is a vector of 21 values, resulting in 41 values in total.} The in total 41 features are extracted solely from the domain name which is to be classified and thus FANCI works completely contextless. FANCI does not support DGA multiclass classification.

\paragraph{\textbf{Pleiades}}
Antonakakis et al.~\cite{antonakakis_throwaway_2012} presented Pleiades, a system which is able to detect machines within a monitored network that are compromised with DGA-based malware. First, unsupervised clustering is used in order to group similar domains of multiple machines within the monitored network. Second, using supervised learning of alternating decision trees a set of NXDs for a given host is labeled to a specific cluster that represents a DGA. Lastly, for each of these clusters a Hidden Markov Model is trained and used for finding single active domain names which are likely to be C2 domains of a particular DGA.

\paragraph{\textbf{M-Endgame}}
Woodbridge et al.~\cite{woodbridge_predicting_2016} propose two RNN-based classifiers, one for DGA binary detection and one for the DGA multiclass classification task. Both classifiers consist of an embedding layer, a long short-term memory (LSTM)~\cite{hochreiter_long_1997} layer containing 128 hidden cells with hyperbolic tangent activation, and a final output layer. In case of binary classification, the final output layer incorporates a single neuron with sigmoid activation which yields a confidence score between zero and one for every input domain. For multiclass classification, the final output layer includes as many neurons as DGA families are included during the training of the classifier. Using the softmax activation function it is determined whether the domain is benign or, in the case of a malicious domain, which DGA is most likely to have generated the domain. We refer to the multiclass classification model as M-Endgame in the following.

\paragraph{\textbf{M-Endgame.MI}}
Woodbridge et al.~\cite{woodbridge_predicting_2016} identified that their proposed M-Endgame model is prone to class imbalances which limit the correct attribution of domains from weakly represented DGAs. The authors tried to cope with this issue by clustering the 30 investigated DGAs into eleven super families. While this increased the averaged classification performance, it removes the possibility of targeted remediation measures. Tran et al.~\cite{tran_lstm_2018} addressed this issue by introducing class weights $C_{i}$ for every class $i$ which control the extent of the weight updates during the training of the M-Endgame model. By weighting the loss function, misclassified samples of class $i$ are penalized with a factor of $C_{i}$ instead of $1$. A greater $C_{i}$ thus forces the classifier to emphasize more on the class $i$. The proposed class weights by Tran et al. are defined as follows:

\begin{displaymath}
	C_{i} = \Big ( \frac{total\ number\ of\ samples}{number\ of\ samples\ in\ class\ i} \Big )^{\gamma}
\end{displaymath}
\\
The parameter $\gamma$ regulates the magnitude of the rebalancing which is to be applied. When $\gamma = 0$ is set, the model behaves cost-insensitive. Setting $\gamma = 1$ causes the classifier to treat every class equally regardless of how many training samples per class are available. Tran et al. evaluated their approach against RUSBoost \cite{seiffert_rusboost_2009} which was shown in \cite{galar_review_2011} to be one of the best performing and least complex approaches in order to cope with class imbalances compared to several bagging-, boosting-, and hybrid-based approaches. They empirically determined $\gamma=0.3$ to work well for M-Endgame and for the DGA multiclass classification task and demonstrated that their approach achieves better results than RUSBoost. In the following, we refer to the adapted cost-sensitive model as M-Endgame.MI.

\paragraph{\textbf{NYU}}
Zhang et al.~\cite{zhang_characterlevel_2015} proposed a CNN-based classifier based on six stacked 1-dimensional convolutional layers for different natural language text classification tasks. While this model was successfully applied to full-fledged natural texts such as news articles or reviews, the model tends to overfit domain names due to their typical small length and missing grammatics. Thus, for the DGA binary classification task, Yu et al.~\cite{yu_character_2018} adapted the model by reducing the total amount of convolutional layers to two and the number of their filters to 128.

\paragraph{\textbf{eXpose}}
Saxe and Berlin~\cite{saxe_expose_2017} proposed eXpose, a CNN-based classifier for detecting malicious URLs, file paths and registry keys. The main difference of this model compared to NYU lies in the usage of the CNN layers. While the NYU model makes use of two stacked CNN layers, the eXpose model uses four parallel CNN layers. This model was successfully applied to DGA binary classification in  \cite{yu_character_2018}.

\paragraph{\textbf{M-ResNet}}
Drichel et al.~\cite{drichel_analyzing_2020} proposed a binary and a multiclass DGA classifier based on ResNets. ResNets introduce skip connections between convolutional layers which build up residual blocks and allow the gradient to bypass layers unchanged during the training of a classifier. Thereby, the vanishing gradient problem can be mitigated. For the binary classification task, the authors propose a classifier consisting of a single residual block with 128 filters per convolutional layer. The proposed multiclass classifier M-ResNet  possesses a more complex architecture of eleven residual blocks and 256 filters per layer.

\subsection{Selected State-of-the-Art Classifiers}
\label{sec:selected_sota_classifiers}
In this section, we provide the rationales behind our choice of state-of-the-art classifiers to comparatively evaluate against. We selected several approaches in order to cover different types of machine learning techniques, i.e. feature-, RNN-, CNN-, and ResNet-based approaches. Moreover, we explain every modification we did in order to enable a fair comparative evaluation.

For a feature-based approach, we adapted FANCI with its original features developed for the binary task to a multiclass classifier. While the RF-based implementation is inherently capable of multiclass classification, the SVM approach requires modifications. By reducing the problem of multiclass classification to multiple binary classification problems it is possible to enable multiclass classification support for the SVM implementation. Here, we either use multiple one-vs.-one (OvO) or one-vs.-rest (OvR) classifiers. We refer to the multiclass enabled SVM approaches as \textbf{\mbox{M-FANCI-SVM-OvO}} and \textbf{\mbox{M-FANCI-SVM-OvR}} in the following. Additionally to the multiclass enabled \textbf{\mbox{M-FANCI-RF}} model we also evaluate \textbf{\mbox{M-FANCI-RF-OvO}} and \textbf{\mbox{M-FANCI-RF-OvR}}.

We do not evaluate Pleiades as it heavily depends on extensive tracking of DNS traffic which is required in order to correlate information of groups of DNS queries or responses. Moreover, the proposed DGA multiclass classifier requires a set of NXDs from a single host as input because it is not able to reliably extract features from a single domain name \cite{antonakakis_throwaway_2012}.

For deep learning based approaches, we chose \textbf{\mbox{M-Endgame}} (RNN-based) and \textbf{\mbox{M-ResNet}} (ResNet-based). Further, we adapted the NYU classifier (CNN-based) to a multiclass model (\textbf{\mbox{M-NYU}}) by exchanging the last output layer similar to the M-Endgame model as proposed in \cite{drichel_analyzing_2020}. Another representative for a CNN-based approach is eXpose. We decided to evaluate the NYU model instead of eXpose in more detail, as prior evaluations \cite{yu_character_2018} suggest that there is only little difference in terms of accuracy between these models but the NYU model is faster to train and needs less time for classification.

Finally, we also include cost-sensitive models as proposed by Tran et al.~\cite{tran_lstm_2018}. In \cite{drichel_making_2020} it was shown that besides the M-Endgame model also other neural network classifiers benefit from class weighting. We thus include for every chosen deep learning based classifier also a cost-sensitive variant for our evaluation. We denote the cost-sensitive models by an ending \textbf{.MI} in the following.

% !TEX root = ../paper.tex
\section{Evaluation Overview}
\label{sec:evaluation_overview}
In this section, we provide an overview of the used data sets as well as our experimental setup which includes our used evaluation methodology.

\subsection{Data Sets}
\label{sec:datasets}

First,  we describe our used data sources and subsequently we explain how we created two distinct data sets, one for feature engineering and selection (\textbf{Set\textsubscript{Selection}}), and one for the final comparative evaluation (\textbf{Set\textsubscript{Evaluation}}). We obtain domain names from two separate sources, one for malicious and one for benign samples.

\subsubsection{Malicious Data: DGArchive}
We use the open-source threat intelligence feed of DGArchive~\cite{plohmann_comprehensive_2016} as source for malicious domain names. DGArchive contains domains generated by reimplementations of DGAs and known seeds. For our evaluation, we use all available samples up to January 1\textsuperscript{st}, 2020. Overall DGArchive provides us with approximately 115 million unique domains generated by 92 different DGAs.

\subsubsection{Benign Data: University Network}
We obtain benign NXDs from the central DNS resolver of the campus network of RWTH Aachen University. This network assimilates several academic and administrative networks, student residences' networks, networks from a university hospital, and eduroam~\cite{eduroam}. We captured a one-month recording of September 2019 from this network that includes approximately 26 million unique NXDs.

\subsubsection{Data Set Creation}
Before creating any data set, we perform a simple data sanitization step in which we convert all domains to lower case and remove all duplicates as well as invalid domains (according to \cite{mockapetris_domain_1987}). Casting the domain names to lower case eases the training of the deep learning based classifiers as they operate on character-level. This data sanitization has no impact on the classification or the name resolution of the DNS as it operates case-insensitive. Moreover, invalid domains can be removed as  their name resolution would fail anyways. Finally,  we filter our benign labeled domain names against all samples of DGArchive and remove all known malicious AGDs  to clean our data as far as possible. In order to obtain meaningful results for our feature engineering, feature selection, as well as for the comparative evaluation, we require that for every included DGA at least 10 unique samples are available. We thus eliminate the samples of \textit{Dnsbenchmark} and \textit{Randomloader}, for which only 5 samples per DGA family are known, from our data. 

Since we aim for diverse data sets, we randomly draw for every remaining DGA in DGArchive at most 20,000 samples. We take all available domain names for DGAs for which less than 20,000 samples are known. Additionally, we draw 20,000 random samples from our source for benign data. Thereafter, we split the selected data evenly across all class labels into two disjoint data sets. The first set, \textbf{Set\textsubscript{Selection}}, is used in the context of feature engineering and selection during the development of our proposed classifier. The second set, \textbf{Set\textsubscript{Evaluation}}, is only used for the final comparative evaluation. Each of these data sets comprises approximately 500,000 samples of 91 different classes including the benign class.

\subsection{Experimental Setup}
We use the following software packages for our experiments: Python 3.8.5, scikit-learn 0.23.2, TensorFlow 2.3.0, Keras 2.4.0, CUDA 10.1, and cuDNN 7.6.5. All deep learning models are executed on a NVIDIA Tesla V100 GPU while the feature-based approaches are executed on 48 CPU cores of two Intel Xeon Platinum 8160 processors@2.1GHz.

Our evaluation is split into two parts (Section~\ref{sec:contribution} and Section~\ref{sec:evaluation}).

In Section~\ref{sec:contribution}, we present evaluations conducted during the development of our classifier. We make use of samples included in {Set\textsubscript{Selection}} in order to engineer and select well performing feature sets. Here, we additionally perform the hyperparameter optimization. The results of the evaluations performed in this section are several promising combinations of feature sets and hyperparameters (configurations) which will be analyzed subsequently.

In Section~\ref{sec:evaluation}, we compare our best performing classifier configurations with the various state-of-the-art classifiers proposed in related work using the samples included in {Set\textsubscript{Evaluation}}. Additionally, we measure the classifiers' training and classification speed in order to assess their real-time capability.

For every evaluation we present in this paper, we perform five repetitions of a five-fold cross validation stratified over the included classes within the respective set. Thus, in every fold, the samples of each class are split into 80\% training and 20\% testing samples.  For the deep learning classifiers we additionally split 5\% from the training samples for a holdout set which is used to assess the performance of the classifiers during training. We train all deep learning models as long as they are improving on the holdout set. After five epochs without improvement we stop the training and evaluate the best model on the test samples.

In order to evaluate and compare the different classifiers we primarily use the f1-score which is defined as the harmonic mean of the precision and the recall. The precision measures the fraction of true positives among those samples that are labeled as positive by a classifier. The recall, on the other hand, equals the true positive rate and thus measures the proportion of positives that are correctly identified by a classifier. We calculate these metrics for every class included in our evaluation and afterwards average all class scores to assess the overall performance of a classifier. By using this macro-averaging we value each class with the same level of importance despite the actual number of available samples per class varying.

% !TEX root = ../paper.tex
\section{EXPLAIN}
\label{sec:contribution}

The engineering of features requires much more effort compared to the usage of deep learning classifiers where all important information has simply to be encoded and provided to the model, then the model learns the relevant features on its own in an end-to-end fashion. Moreover, after the feature engineering the best combination of features has to be selected. The combination of several engineered features might contain mutual information which could render single features useless for the classification. These features should be removed as their extraction from the raw data might require significant processing time which could have a negative impact on the real-time capability of a classifier. For performing the complex process of feature selection a multitude of feature filtering and ranking techniques (e.g. {\cite{guyon2002gene, gilles2013understanding, breiman2001random, kononenko1997overcoming, urbanowicz2018benchmarking, kira1992feature}) have been suggested in the past. Lastly, in the development process of a feature-based classifier a huge amount of hyperparameters has to be optimized. All in all, for a promising feature-based classifier these three steps (i.e. feature engineering, feature selection, and hyperparameter optimization) have to be performed which is not a trivial task.

During the development of EXPLAIN we investigated RFs and SVMs. In our experiments, our RF variants outperformed our SVM approaches in training time as well as in classification performance. We thus focus on the development of a RF-based implementation of a feature-based multiclass DGA classifier in the following. 

\subsection{Feature Engineering \& Selection}
\label{sec:feature_selection}

\begin{table*}[!t]
	\caption{Newly Developed Features}
	\label{tab:new_features}
	\centering
	\resizebox{\linewidth}{!}{
		\begin{tabular}{llll}
			\toprule
			\textbf{Feature} & \textbf{\#} & \textbf{Type} & \textbf{Goal/Purpose} \\
			\midrule
			subdomains-digit-sum         & 1  & linguistic  & distinguish families with different character distributions \\
			\{...\}-character-ratio      & 4  & linguistic  & distinguish families with different character distributions \\
			alphabet-\{...\}             & 37 & linguistic  & distinguish families with different character distributions and alphabets \\
			adjacent-duplicates-ratio    & 1  & linguistic  & distinguish arithmetic- and wordlist-based families from hex-based families \\
			\{...\}-max-streak-length    & 6  & linguistic  & distinguish arithmetic-, hex- and wordlist-based families \\
			first-character-pair         & 1  & linguistic  & detect families with constant prefixes (e.g. \textit{Xxhex} with ``xx'') \\
			syllable-count               & 1  & linguistic  & distinguish families with different levels of readability in their AGDs \\
			weighted-steaks              & 1  & linguistic  & distinguish same length domains with different positioning and quantity of consonants and decimal digits \\
			inverse-hamming-distance     & 1  & linguistic  & for randomness assessment \\
			\{...\}-digit-edge-distance  & 2  & linguistic  & detect domains with digits at constant relative positions and for randomness assessment \\
			suffix-digit-sum             & 1  & linguistic  & distinguish families with different sets of utilized suffixes \\
			suffix-standard-deviation    & 1  & linguistic  & distinguish families with different sets of utilized suffixes \\
			suffix-length                & 1  & structural  & distinguish families with different sets of utilized suffixes \\
			suffix-dns-level             & 1  & structural  & distinguish families with different sets of utilized suffixes \\
			second-level-length          & 1  & structural  & detect families using fixed lengths for their subdomains \\
			subdomains-length            & 1  & structural  & detect families using fixed lengths for their subdomains \\
			second-level-repeated-prefix & 1  & structural  & detect domains originating from misconfigured software \\
			subdomains-contain-suffix    & 1  & structural  & detect domains originating from misconfigured software and typing errors \\
			\{1, 2, 3\}-gram-\{...\}     & 15 & statistical & distinguish families with different character distributions and for randomness assessment \\
			zlib-bits-compression-ratio  & 1  & statistical & distinguish families with different character distributions and for randomness assessment \\
			bits-entropy                 & 1  & statistical & distinguish families with different character distributions and for randomness assessment \\
			\{...\}-test[-unicode]       & 14 & statistical & distinguish different underlying pseudo-random number generators \\
			\bottomrule
		\end{tabular}
	}
\end{table*}

In this work, we study 136 different features which we gathered or adapted from related work (mainly from  \cite{schuppen_fanci_2018,antonakakis_throwaway_2012,schiavoni_phoenix_2014,plohmann_comprehensive_2016}) and developed by our own through analyzing the samples of the different DGAs and benign samples contained in {Set\textsubscript{Selection}}. As we target a contextless classifier due to privacy considerations we only focus on features that can be extracted from a single domain name. We divide the 136 features into 64 linguistic, 17 structural, and 55 statistical features. Note, we only use samples from {Set\textsubscript{Selection}} for feature engineering and selection. The final comparative evaluation will be performed on samples from the disjunctive set {Set\textsubscript{Evaluation}}.

We provide a list of the developed features including their total amount and a brief description of their purpose in Table~\ref{tab:new_features}. For instance, simple features attempt to distinguish DGA families based on the suffixes used in their generated domains because different families use different sets of suffixes. Other features try to separate DGA families based on character distributions as, for example, wordlist-based families do have different consonant and vowel distribution in contrast to arithmetic- and hex-based families. Additionally, we introduce novel features to discriminate different underlying pseudo-random number generators used by the DGA families for domain generation. A detailed description for each individual feature can be found in the source code.

After feature engineering we perform feature selection to reduce the computational complexity for training and classification of our classifier and to enhance its overall classification performance. A variety of different feature selection methods have been proposed in the past, all having their advantages and disadvantages. In this work, we thus make use of different filter and wrapper methods to determine valuable features as there is no best technique. 

Filtering methods leverage proxy measures to assess the importance of features. The main advantage of filter methods are that they are computationally lightweight, scalable, and independent of the underlying learning algorithm. Common measures are the variance, mutual information, chi-square test, ANOVA F-test, and Relief-based algorithms~\cite{kira1992feature}.

The variance as well as the mutual information of a variable with a target label can be used as a proxy to measure the amount of information of a feature. Features that contain little information can be filtered out because they contribute only insignificantly to the classification.

The chi-square test measures the dependence between a non-negative categorical feature and the target label which can thus be used to remove features that are independent of a class and therefore irrelevant for the classification. In case of numerical features the ANOVA F-test should be used instead. As we intent to utilize a mix of different categorical and numerical features we do not make use of these filtering techniques. 

\begin{table*}[!t]
	\caption{Feature Selection Analysis}
	\label{tab:feature_selection_analysis}
	\centering
	\resizebox{\linewidth}{!}{
		\begin{tabular}{lccccccc}
			\toprule
			\textbf{Feature Set} & \textbf{\#} & \textbf{F1-score} & \textbf{Precision} & \textbf{Recall} & $\frac{\textbf{Training}\ \textbf{Time}}{\textbf{Classifier}} \textbf{[s]}$ & $\frac{\textbf{Feature}\ \textbf{Extraction}\ \textbf{Time}}{\textbf{Sample}} \textbf{[}\boldsymbol\mu \textbf{s]}$ & $\frac{\textbf{Inference}\ \textbf{Time}}{\textbf{Sample}} \textbf{[}\boldsymbol\mu \textbf{s]}$\\
			\midrule
			RFE-MDI & \phantom{0}52 & 0.74290 & 0.78390 & 0.73592 & 195 & 156 & 17 \\
			RFE-PI & \phantom{0}28 & 0.75504 & 0.78528 & 0.74934 & 100 & 106 & 16 \\
			ReliefF & \phantom{0}41 & 0.71707 & 0.74192 & 0.71267 & 184 & 198 & 18 \\
			MultiSURF & \phantom{0}59 & 0.72946 & 0.77833 & 0.72353 & 192 & 241 & 17 \\
			\midrule
			All features & 136 & 0.72806 & 0.77131 & 0.72114 & 320 & 695 & 17 \\
			Intersection & \phantom{0}11 & 0.64527 & 0.67936 & 0.64778 & \phantom{0}27 & \phantom{0}85 & 16 \\
			Union & \phantom{0}76 & 0.73352 & 0.77842 & 0.72662 & 264 & 276 & 17 \\
			Union-Spearman & \phantom{0}64 & 0.74654 & 0.79204 & 0.73836 & 225 & 239 & 17 \\
			\bottomrule
		\end{tabular}
	}
\end{table*}

Relief-based algorithms~\cite{kira1992feature} compute an importance score for every feature based on differences in feature values of nearest neighbor instance pairs. The advantages of Relief-based algorithms is that they run in low-order polynomial time, are not sensitive to feature interactions, and are robust against noise. However, their weaknesses are that they do not discriminate redundant features and that they might be fooled by low numbers of training instances. ReliefF~\cite{kononenko1997overcoming} is the most commonly used Relief-based algorithm \cite{urbanowicz2018benchmarking}. Another efficient Relief-based algorithm is MultiSURF~\cite{urbanowicz2018benchmarking}. It can be used for the sake of simplicity or when computing resources are limited, since there are no execution parameters to be optimized. For our feature selection, we utilize ReliefF and MultiSURF to select the better than average features according to their computed feature importances.

Wrapper methods are, in contrast to filter methods, more computationally intensive as they are not independent of the underlying learning algorithm. For each feature subset a new model is trained and evaluated which allows for finding the best performing feature set for a particular model and evaluation set.

A commonly used wrapper method is recursive feature elimination (RFE)~\cite{guyon2002gene}. RFE recursively estimates the feature importances based on a feature ranking method and removes beginning from the whole feature set the least important feature. In each iteration of RFE the classification performance of the current feature set is estimated using a hold-out set. Thus, using RFE it is possible to determine the best feature set for a given model and evaluation set. In this work, we use the cross-validated variant of RFE included in scikit-learn~\cite{scikit-learn} using the Mean Decrease Impurity (MDI)~\cite{gilles2013understanding} and the Permutation Importance (PI)~\cite{breiman2001random} as feature ranking methods. 

MDI measures for every feature the average gain of purity by splits using the corresponding feature within the trees in the forest. The advantage of MDI is that once the model is trained the MDI for each feature can be calculated without the need of further model executions (e.g. evaluating a hold-out set). However, this property is at the same time a weakness of MDI as the feature importances are only derived from statistics of the training data set and thus this measure does not indicate which features would be most important for good predictions on a hold-out set. Further, MDI might overestimate the importances of high cardinality features.

PI, on the other hand, does not suffer from these issues. After a model is trained, the PI can be measured for every feature by calculating the increase of the model's prediction error after permuting the feature values included in a validation set. If the model error increases (compared to the model error measured on the validation set containing the unshuffled feature values), the feature is important as the model uses it for correct predictions. The feature is not important if the model error stays unchanged as the model does not rely on the corresponding feature for classification.

To select valuable features, we make use of RFE with MDI as well as RFE with PI.

\paragraph{\textbf{Results}}
For our feature selection, we first exclude ill-defined features with zero variance or zero mutual information with the target label since they do not contain any information that a classifier could use to distinguish samples. Thereby we remove three of the 136 investigated features. Thereafter, we derive four different feature sets using an RF classifier with default hyperparameters (set by scikit-learn~\cite{scikit-learn}) and the previously introduced feature selection methods: \textbf{ReliefF}, \textbf{MultiSURF}, \textbf{RFE-MDI}, and \textbf{RFE-PI}.

As there is no best feature selection method, we additionally combine all sets into an \textbf{Intersection} set (by calculating the intersection of all sets) and a \textbf{Union} set (by taking the union of all sets). Additionally, we remove multicollinear features within the Union set. Removing such features could improve the classifier's training and classification time without decreasing its classification performance since the classifier can obtain the same information from correlating features. The removal of such features thus reduces the computational burden of a classifier. In order to achieve this, we perform hierarchical clustering on the features' Spearman rank-order correlation coefficients \cite{spearman1961proof}, where the coefficients measure the monotonicity of the relationships between different features. 

We provide a heatmap of correlating features contained in the Union set in Fig.~\ref{fig:correlation_dendogram_matrix} in the Appendix. The darker a cell within the figure, the more the features, which are depicted on the x- and y-axis, correlate positively. It can be seen, that while the features in the left upper part of the heatmap are less correlated to each other, features within the right lower part build several clusters.

In order to remove the correlating features we calculate a cut-off threshold targeting the preferred number of remaining clusters. For better understanding, we provide the corresponding dendrogram in Fig.~\ref{fig:correlation_dendogram_matrix} on top of the heatmap including the plot of the cut-off threshold. The y-axis is a measure of closeness of the different clusters. In our case the calculated threshold equals approximately 0.39. The features which are clustered under the threshold line are collapsed by choosing the feature with the highest MDI. Through this process we generate the additional feature set \textbf{Union-Spearman}.

The upper part of Table \ref{tab:feature_selection_analysis} displays the amount of selected features per individual selection method as well as evaluation results obtained by classifying the samples of {Set\textsubscript{Selection}}. In the lower part, we additionally include results of the different feature set combinations as well as an evaluation using all 136 features for comparison.

The best evaluation results on {Set\textsubscript{Selection}} are achieved with RFE-PI (f1-score of $75.504\%$). This is remarkable since only 28 of the 136 features are used. The only feature set which contains less features is the Intersection set with eleven features in total. However, the f1-score for this set is with $64.527\%$ the worst. The training and feature extraction speed is the best for RFE-PI when the Intersection set is ignored. All feature selections, except for ReliefF and Intersection, improve the classification performance compared to the classifier that uses all features. The removal of twelve multicollinear features contained in the Union set (yielding Union-Spearman) increases the f1-score by more than $1\%$. The required inference time per sample does not vary much between all feature sets.

Since it cannot be ruled out that some feature sets might perform well on the utilized data set due to overfitting, we consider for our further development the best individual feature set (RFE-PI) as well as the Union and Union-Spearman feature set combinations. We thus perform individual hyperparameter optimizations for all of these three feature selections.

\subsection{Selected Features}
Here, we only present the features we have selected. The full list of investigated features as well as a description for each individual feature can be found in the publicly available source code.

All selected features can be separated into three different groups: linguistic, structural, and statistical features. The first category of features analyze the presence or absence of common linguistic patterns. For instance, features of this category evaluate whether a domain name contains digits or compute the vowel ratio. Structural features, on the other hand, investigate structural properties of a domain name such as the domain length. The last category contains features which capture statistical properties such as the frequency distribution of certain n-grams or the entropy.

In Table~\ref{tab:all_features} in the Appendix, we provide an overview of all features selected by the different selection methods. For every feature, we mark the membership to a corresponding feature set and present extracted feature values for two sample domains, $d_0$ and $d_1$. We define a feature as a function $\mathcal{F}$ of a sample $d$. Thus $\mathcal{F}(d)$ denotes the extracted feature. In our example, $d_0 =$ \textit{iee-security.org} represents a benign NXD caused by a typing error of \textit{ieee-security.org} while \mbox{$d_1 =$ \textit{mwkwhvkdpp.info}} is a malicious NXD generated by the \textit{Conficker} DGA.

Note, 60 out of the 76 features from the Union set are newly developed indicating the need of new features for the DGA multiclass classification task. In Section~\ref{sec:evaluation1}, we assess our selected feature sets against the features proposed in related work by  performing a comparative evaluation including the feature-based approach FANCI~\cite{schuppen_fanci_2018}.

\subsection{Hyperparameter Optimization}
\label{sec:hyperparameter_optimization}
The exhaustive grid search is one of the most used hyperparameter optimization strategy\cite{bergstra2012random}. It generates candidates from a grid of parameter values and evaluates each in order to determine the optimal hyperparameters. As every possible hyperparameter combination for the defined values is evaluated, this brute-force approach is computationally expensive. Random search~\cite{bergstra2012random} can be used in order to reduce computational costs by performing a certain number of randomly chosen trials over the hyperparameter space. Obviously, the parameter values that are to be investigated in a grid search can be reduced in order to decrease the computational costs. However, by a more coarse grid search it is more probable to miss well performing hyperparameter combinations which might be found by a random search. The amount of trials plays a crucial role in finding well performing hyperparameters in a random search. Practically, often 60 trials are used because the maximum of 60 random observations lies within the top 5\% of the true maximum with a probability of 95\%. However, this only holds true if the close-to-optimal region of the hyperparameters occupies at least 5\% of the whole grid surface. Another optimization strategy to tackle the computational costs is the Bayesian search~\cite{mockus1978application}. The Bayesian search is a sequential process which takes information of the previously evaluated hyperparameter combinations into account in order to choose the next set of hyperparameters for evaluation. The downside of the Bayesian search is that it is not parallelizable since the next search always depends on the results of the previous searches.

In this work, we thus choose to utilize random search for hyperparameter optimization. Thereby, we do not require as much computational resources as for an exhaustive grid search but are able to parallelize the optimization. As we do not know how much space the close-to-optimal region of the hyperparameters occupies in our case, we double the recommended amount of random trials to 120 in order to find well performing hyperparameters.

As stated in Section~\ref{sec:feature_selection}, we consider the following feature sets for our hyperparameter optimization: RFE-PI, Union, and Union-Spearman. For each feature set, we run two hyperparameter optimizations, one using the random forest implementation which is inherently capable of multiclass classification (RF) and one using an one-vs.-rest variant (OvR). We discard all one-vs.one (OvO) variants due to their slow classification speed although their classification performance might be slightly better. In detail, we optimize the following hyperparameters of the random forest implementation of scikit-learn~\cite{scikit-learn}: \textit{n\textunderscore estimators}, \textit{criterion}, \textit{max\textunderscore depth}, \textit{max\textunderscore features}, \textit{bootstrap}, and \textit{class\textunderscore weight}.

\paragraph{\textbf{Results}} For all three investigated feature sets (RFE-PI, Union, Union-Spearman) we obtain best results on {Set\textsubscript{Selection}} using the OvR implementation. We refer to the three different combinations of chosen hyperparameters and feature sets as \textbf{EXPLAIN-OvR\textsubscript{RFE-PI}}, \textbf{EXPLAIN-OvR\textsubscript{Union}}, and \textbf{EXPLAIN-OvR\textsubscript{Union-Spearman}} in the following. We note, that the OvR variants require more training and classification time than the RF variants. Thus, we additionally include the fastest RF implementation to the classifier configurations for comparison. The fastest model makes use of the RFE-PI feature set. We refer to this model as \textbf{EXPLAIN-RF\textsubscript{RFE-PI}} in the following. For reproducibility we provide the specific values for each hyperparameter and for every model in the source code.

% !TEX root = ../paper.tex
\section{Comparative Evaluation}
\label{sec:evaluation}

In Section~\ref{sec:evaluation1}, we first present the results of our comparative evaluation. Thereafter, in Section~\ref{sec:evaluation2}, we examine the training and classification speed of the various classifiers in order to assess their real-time capability.

\subsection{Classification Performance}
\label{sec:evaluation1}

\begin{table}[!t]
	\caption{Multiclass Classification Results}
	\label{tab:mcc_results}
	\centering
%	\small
	\resizebox{\columnwidth}{!}{
		\begin{tabular}{lccl}
			\toprule
			\textbf{Classifier} & \textbf{F1-score} & \textbf{Precision} & \textbf{Recall} \\
			\midrule
			M-FANCI-RF & 0.56808 & 0.58680 & 0.57805 \\
			M-FANCI-RF-OvO & 0.57097 & 0.59210 & 0.58092 \\
			M-FANCI-RF-OvR & 0.56907 & 0.58873 & 0.57852 \\
			M-FANCI-SVM-OvO & 0.50320 & 0.55289 & 0.51028 \\
			M-FANCI-SVM-OvR & 0.35113 & 0.38483 & 0.37827 \\
			M-Endgame & 0.74641 & 0.76731 & 0.74327 \\
			M-Endgame.MI & 0.75287 & 0.77100 & 0.75351 \\
			M-NYU & 0.75447 & 0.79080 & 0.74648 \\
			M-NYU.MI & 0.78069 & 0.80698 & 0.78038 \\
			M-ResNet & 0.79574 & \textbf{0.81915} & 0.79224 \\
			M-ResNet.MI & \textbf{0.80361} & 0.81435 & \textbf{0.81036} \\
			\midrule
			EXPLAIN-RF\textsubscript{RFE-PI} & 0.76114 & 0.77862 & 0.75982 \\
			EXPLAIN-OvR\textsubscript{RFE-PI} & 0.76883 & 0.79245 & 0.76624 \\
			EXPLAIN-OvR\textsubscript{Union} & \textbf{0.78554} & \textbf{0.81631} & \textbf{0.77955} \\
			EXPLAIN-OvR\textsubscript{Union-Spearman} & 0.78046 & 0.81541 & 0.77540 \\
			\bottomrule
		\end{tabular}
	}
\end{table}

For our comparative evaluation, we compare the selected state-of-the-art classifiers (Section~\ref{sec:selected_sota_classifiers}) with our developed EXPLAIN classifier configurations (Section~\ref{sec:hyperparameter_optimization}) using samples of Set\textsubscript{Evaluation}. 

We summarize the results of this evaluation in Table~\ref{tab:mcc_results}. The ResNet-based approaches achieve the best results. Our EXPLAIN-OvR\textsubscript{Union} configuration is among our classifiers the best and accomplishes an f1-score of $78.554\%$. By this means, it is the next best classifier after the ResNet-based approaches. The M-NYU.MI model achieves comparable results. An f1-score of $78.554\%$ might appear rather low but devices infected with DGA-based malware will typically generate multiple AGDs per day. Thus real-world counteractions would not have to be triggered based on a single classification but rather on the fact that multiple AGDs were attributed to the same DGA. 

All deep learning models benefit from class weighting. The NYU model profits the most and improves its f1-score by over $2.6\%$. The adapted feature-based approaches of related work perform poorly. Our best EXPLAIN configuration is by over $21.45\%$ in f1-score better than the best FANCI-based approach. EXPLAIN-OvR\textsubscript{Union-Spearman} is slightly worse than EXPLAIN-OvR\textsubscript{Union}. EXPLAIN-OvR\textsubscript{RFE-PI} and EXPLAIN-RF\textsubscript{RFE-PI} achieve with an f1-score of approximately $76\%$ slightly worse results compared to the EXPLAIN configurations that are based on the Union feature set. However, both of these classifiers are better than all state-of-the-art classifiers except for M-NYU.MI and the ResNet-based approaches.

To better visualize the classification performance and to compare our best classifier configuration (\mbox{EXPLAIN-OvR\textsubscript{Union}}) with the best classifier proposed in related work (M-ResNet.MI), we present a combined confusion matrix in Fig.~\ref{fig:combined_cm}.

The combined confusion matrix shows for both classifiers the relative amount of samples belonging to classes displayed on the vertical axis that are labeled as classes shown on the horizontal axis. For every combination of true and predicted label, space in form of a square is reserved within the figure. Each square is halved into two triangles where the upper left triangle is dedicated to the samples classified by EXPLAIN-OvR\textsubscript{Union} and the lower right triangle visualizes the classification performance of M-ResNet.MI. The individual achieved scores for both classifiers and every class are encoded within the respective triangles as shades of either blue (\mbox{EXPLAIN-OvR\textsubscript{Union}}) or red (M-ResNet.MI). An f1-score of 0\% is encoded as a transparent triangle and 100\% is represented by a fully opaque triangle. A perfect classifier would produce only opaque triangles on the identity matrix diagonal. The benign class is located at the upper left part of the figure. Thereafter are the DGA families listed in alphabetical order. 

Surprisingly, both classifiers discriminate most DGA families as well as the benign class equally well. Moreover, several DGA families which are almost not recognized by our approach are also not recognized by M-ResNet.MI (e.g. \textit{Dircrypt}, \textit{Goznym}, \textit{Hesperbot}, \textit{Tempedreve}). These results indicate that the ResNet-based approach might learn similar features. However, three DGA families (\textit{Redyms}, \textit{Tempedrevetdd}, and \textit{Xshellghost}) are only detected by M-ResNet.MI. It might be possible to engineer new discriminating features by investigating the samples of these three classes in order to enable the correct detection by our classifier. Lastly, both classifiers tend to attribute samples of related DGA families to a single class (e.g. \textit{Pykspa}-\textit{Pykspa2} and \textit{Vidro}-\textit{Vidrotid}).

All in all, these results show that our EXPLAIN classifiers are able to achieve competitive results while being at the same time, due to the feature-based approach, far more explainable.

\subsection{Training \& Classification Speed}
\label{sec:evaluation2}

In the following, we compare the training and classification times of the classifiers proposed in related work with our proposed classifiers. Note, we acknowledge that the reported times are difficult to compare as the deep learning classifiers are able to take advantage of GPU processing while the feature-based approaches are evaluated on CPUs. However, the main goal of this study is to determine whether it is realistic to utilize the classifiers for real-time classification. Although, the hardware components may be scaled, the relative time difference in training and classification time allow for a comparison within the group of deep learning classifiers and within the group of feature-based approaches.

\begin{table}[!t]
	\caption{Performance Analysis}
	\label{tab:speed}
	\centering
	\resizebox{\columnwidth}{!}{
		\huge
		\begin{tabular}{lcc}
			\toprule
			\textbf{Classifier} & $\frac{\textbf{Training}\ \textbf{Time}}{\textbf{Classifier}} \textbf{[s]}$ & $\frac{\textbf{Classification}\ \textbf{Time}}{\textbf{Sample}} \textbf{[}\boldsymbol\mu \textbf{s]}$ \\
			\midrule
			M-FANCI-RF & \phantom{000}28 & \phantom{000}198 \\
			M-FANCI-RF-OvO & \phantom{0}2528 & \phantom{0}31\hspace{1mm}711 \\
			M-FANCI-RF-OvR & \phantom{0}1423 & \phantom{000}435 \\
			M-FANCI-SVM-OvO & \phantom{00}114 & 285\hspace{1mm}851 \\
			M-FANCI-SVM-OvR & 13\hspace{1mm}042 & \phantom{0}47\hspace{1mm}239 \\
			M-Endgame & \phantom{0}1862 & \phantom{000}151 \\
			M-Endgame.MI & \phantom{0}1846 & \phantom{000}149 \\
			M-NYU & \phantom{00}596 & \phantom{0000}55 \\
			M-NYU.MI & \phantom{00}577 & \phantom{0000}55 \\
			M-ResNet & \phantom{0}1106 & \phantom{000}133 \\
			M-ResNet.MI & \phantom{0}1028 & \phantom{000}134 \\
			\midrule
			EXPLAIN-RF\textsubscript{RFE-PI} & \phantom{00}307 & \phantom{000}128 \\
			EXPLAIN-OvR\textsubscript{RFE-PI} & \phantom{0}2533 & \phantom{000}231 \\
			EXPLAIN-OvR\textsubscript{Union} & \phantom{0}7036 & \phantom{000}534 \\
			EXPLAIN-OvR\textsubscript{Union-Spearman} & \phantom{0}3036 & \phantom{000}358 \\
			\bottomrule
		\end{tabular}
	}
\end{table}

In Table~\ref{tab:speed}, we display the training time per classifiers and the classification time per samples for all investigated classifiers. All times are measured during the comparative evaluation presented in Section~\ref{sec:evaluation1}. The classification time per sample includes for feature-based approaches the feature extraction time and for deep learning based approaches the required time for the input pre-processing (i.e. converting domain names to a sequence of integers).

We ignore the training and classification times for the FANCI-based approaches as they performed poorly. However, for the sake of completeness we list their times within the table. Regarding the group of deep learning based approaches, the NYU models are fastest in training and predicting followed by the ResNet-based approaches. The Endgame models are the slowest. Within the group of our proposed classifiers, \mbox{EXPLAIN-RF\textsubscript{RFE-PI}} is by far the fastest to train and classifies samples similarly fast as the M.ResNet.MI model although it is executed on CPUs.
EXPLAIN-OvR\textsubscript{RFE-PI} is the second fastest \mbox{EXPLAIN} model followed by \mbox{EXPLAIN-OvR\textsubscript{Union-Spearman}}. The twelve additional features included in the Union feature set, compared to Union-Spearman, significantly increase the training as well as the prediction time.

\begin{figure*}[!t]
	\centering
	\includegraphics[width=1.0\linewidth]{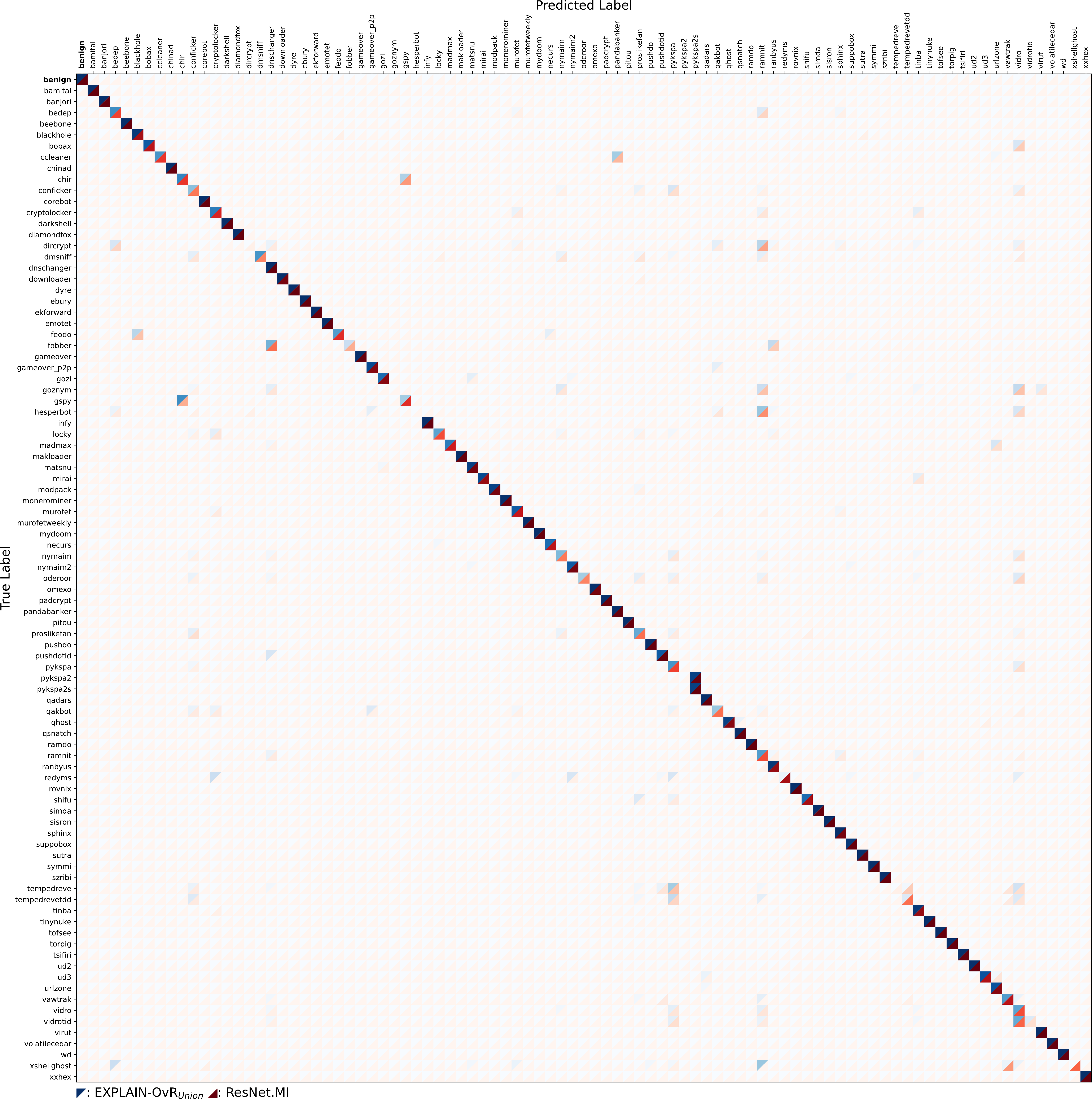}
	\caption{Combined confusion matrix of EXPLAIN-OvR\textsubscript{Union} and M-ResNet.MI.}
	\label{fig:combined_cm}
\end{figure*}

In order to approximate the required classifier's performance for real-time classification, we measure the average amount of occurring NXDs within the university network which was used as source for benign data during our previous experiment. On average, we observed 174 NXD responses per second with an maximum peak of 2325 NXD responses per second. Thus a classifier has at most $430\mu s$ in order to classify a single sample. All classifier under consideration except for \mbox{EXPLAIN-OvR\textsubscript{Union}} are thus able to perform real-time classification. \mbox{EXPLAIN-OvR\textsubscript{Union}} requires $534\mu s$ and thus classifies $1872$ samples per second. However, during the whole one-month recording, the maximum packets per second exceeded only for four consecutive seconds the mark of 1872. Thus for the live classification of a large university network, EXPLAIN-OvR\textsubscript{Union} would only delay a few packets for a few seconds within a whole month. We thus argue that EXPLAIN-OvR\textsubscript{Union} is also well-suited for live classification. Moreover, our proposed EXPLAIN classifiers are highly parallelizable and scale extremely well with the number of CPU cores.

% !TEX root = ../paper.tex
\section{Conclusion \& Discussion}
\label{sec:conclusion}

In this work, we proposed EXPLAIN, a feature-based and contextless DGA multiclass classifier and compared different EXPLAIN configurations with several state-of-the-art classifiers in a unified setting on the same real-world data. 

Our best performing model, EXPLAIN-OvR\textsubscript{Union} uses 76 features and achieves the best f1-score after the ResNet-based approaches. 

EXPLAIN-OvR\textsubscript{RFE-PI} and EXPLAIN-RF\textsubscript{RFE-PI} make use of only 28 features and beat all feature-based approaches proposed in related work by a huge margin. Moreover, they achieve higher f1-scores than the deep learning based approaches: M-Endgame, M-Endgame.MI, and M-NYU. 

Surprisingly, the in detail comparison between \sloppy EXPLAIN-OvR\textsubscript{Union} and M-ResNet.MI indicates that the deep learning classifier might learn very similar features as the ones we have selected for our EXPLAIN classifiers. 

Finally, we analyzed the real-time capability of the various classifiers. All of our proposed classifiers are highly capable of real-time classification. Our fastest proposed model, EXPLAIN-RF\textsubscript{RFE-PI}, is even able to classify 7812 samples per second.

Which EXPLAIN configuration to choose depends on the individual requirements on the classifier. \mbox{EXPLAIN-OvR\textsubscript{Union}} achieves the best classification results using 76 features.
Selecting a configuration that uses the RFE-PI feature set with only 28 features could make it easier to interpret predictions from a model. EXPLAIN-OvR\textsubscript{Union-Spearman} offers a compromise as it uses 64 features but achieves comparable classification results as EXPLAIN-OvR\textsubscript{Union}. The EXPLAIN-RF\textsubscript{RFE-PI} configuration should be chosen over EXPLAIN-OvR\textsubscript{RFE-PI} when the computational resources are limited. Otherwise, for a fast, explainable, and slightly better classifier \mbox{EXPLAIN-OvR\textsubscript{RFE-PI}} should be chosen.

By design our feature-based approach is more explainable compared to the deep learning classifiers proposed in related work as predictions can be traced back to the characteristics of the used of features. In contrast, deep learning classifiers only output a vector of probabilities indicating to which class a particular domain can be attributed to without referring to the actual input.

By proposing a competitive feature-based approach we made a first step towards explainable DGA multiclass classification. Ultimately, a focused comparison of different approaches to DGA multiclass classification with respect to explainability is required. Our work is a necessary prerequisite for a comparative explainability study in which competitively performing feature-based and deep learning based approaches have to be contrasted. In future work it is required to compare the level of explainability provided by our approach with different techniques, such as Lemna~\cite{guo2018lemna} and \mbox{DMM-MEN}~\cite{guo2018explaining}, which try to explain the predictions of deep neural network classifiers. Moreover, recently, a visual analytics system~\cite{becker_interpretable_2020} was proposed which strives to provide understandable interpretations for predictions of deep learning based DGA detection classifiers. This system first clusters the activations of a model's neurons and subsequently leverages decision trees in order to explain the constructed clusters. 

Additionally, future work could analyze the interpretations of deep learning based models for correctly classified samples which, however, are incorrectly classified by our approach. It might be possible to extract used features of deep learning models which enable the correct classification of such samples for our classifier. Thereby, it might be possible to further enhance the performance of our approach.

\begin{acks}
		This project has received funding from the European Union's Horizon 2020 research and innovation programme under grant agreement No 833418. 
		Simulations were performed with computing resources granted by RWTH Aachen University under project rwth0438.
\end{acks}

\bibliographystyle{ACM-Reference-Format}
\bibliography{bibliography}

\appendix
% !TEX root = ../paper.tex
\section{Appendix}
\label{sec:appendix}

In Fig. \ref{fig:correlation_dendogram_matrix}, we support the removal of multicollinear features within the Union set (see Section~\ref{sec:feature_selection}) by visualizing the heatmap of correlating features together with a dendrogram.

In Table \ref{tab:all_features}, we provide an overview of all features selected by the different selection methods including the membership to a corresponding feature set and extracted feature values for two sample domains, $d_0$ and $d_1$. A feature is defined as a function $\mathcal{F}$ of a sample $d$. $\mathcal{F}(d)$ denotes the extracted feature value. In our example, $d_0 =$ \textit{iee-security.org} represents a benign NXD caused by a typing error of \textit{ieee-security.org} and $d_1 =$ \textit{mwkwhvkdpp.info} is a malicious NXD generated by the \textit{Conficker} DGA.

\begin{figure*}[!t]
	\centering
	\includegraphics[width=1.00\linewidth]{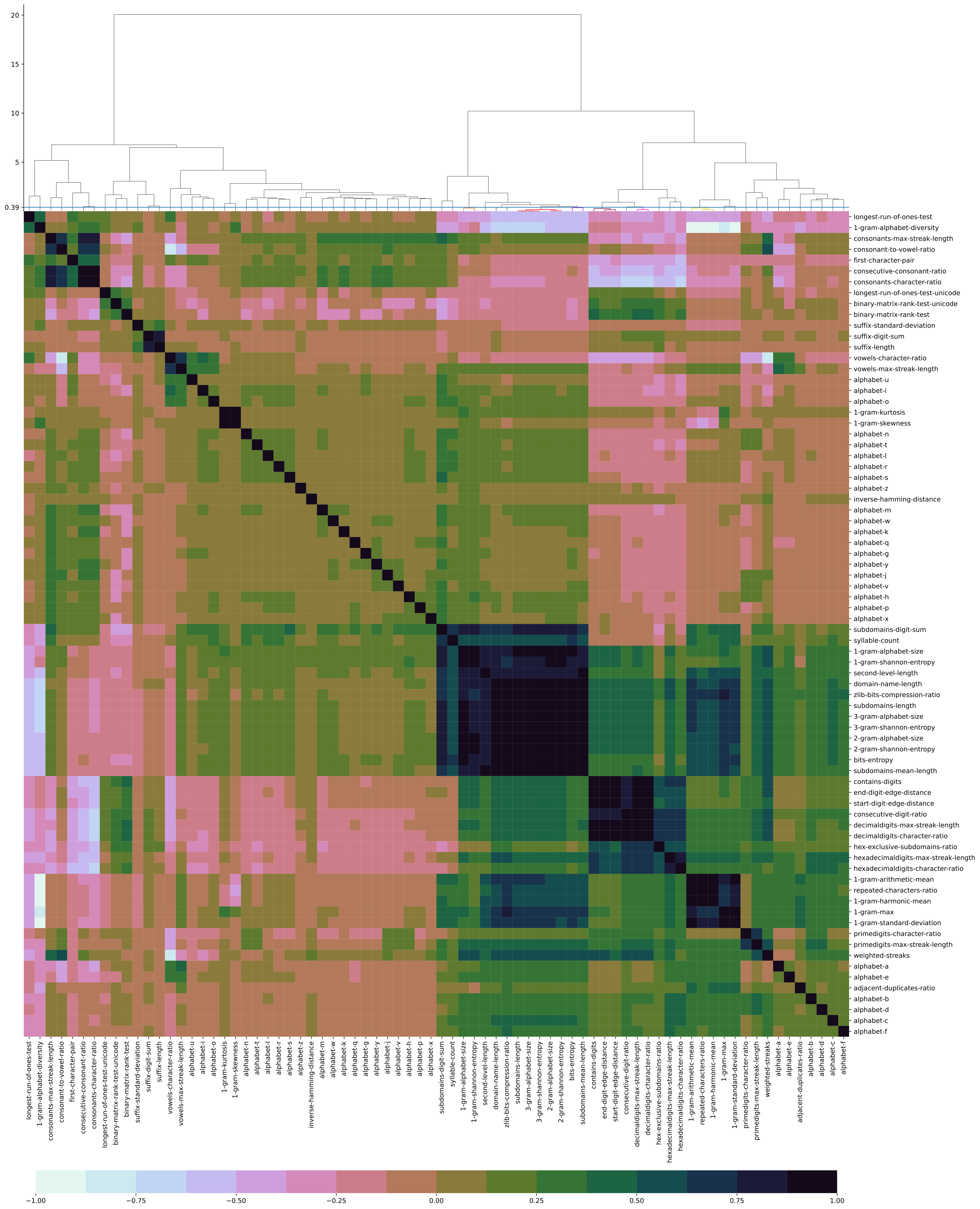}
	\caption{Heatmap and dendrogram of the correlating features of the Union feature set.}
	\label{fig:correlation_dendogram_matrix}
\end{figure*}

% !TEX root = ../paper.tex
\begin{table*}[!t]
    \caption{\textbf{All Selected Features:} $\textbf{d\textsubscript{0} =}$ \textbf{\textit{iee-security.org}}\textbf{,} $\textbf{d\textsubscript{1} =}$ \textbf{\textit{mwkwhvkdpp.info}}}
    \label{tab:all_features}
    \centering
    \tiny
    \resizebox*{\linewidth}{\textheight}{
    \begin{tabular}{rlcccccllrr}
        \toprule
        \textbf{\#} & \textbf{Feature} & \textbf{RFE-MDI} & \textbf{RFE-PI} & \textbf{RELIEFF} & \textbf{MultiSURF} & \textbf{Union-Spearmann} & \textbf{Type} & \textbf{Output} & $\mathcal{F}(d_0)$ & $\mathcal{F}(d_1)$\\
        \midrule
        1           & adjacent-duplicates-ratio& \cmark           &                 &                  & \cmark             & \cmark                    & linguistic    & rational        & 0.08333                     & 0.10000                     \\
        2           & alphabet-a       & \cmark           & \cmark          & \cmark           & \cmark             & \cmark                    & linguistic    & integer         & 0                           & 0                           \\
        3           & alphabet-b       &                  & \cmark          &                  & \cmark             & \cmark                    & linguistic    & integer         & 0                           & 0                           \\
        4           & alphabet-c       &                  &                 &                  & \cmark             & \cmark                    & linguistic    & integer         & 1                           & 0                           \\
        5           & alphabet-d       &                  &                 & \cmark           & \cmark             & \cmark                    & linguistic    & integer         & 0                           & 1                           \\
        6           & alphabet-e       &                  &                 & \cmark           & \cmark             & \cmark                    & linguistic    & integer         & 3                           & 0                           \\
        7           & alphabet-f       &                  &                 &                  & \cmark             & \cmark                    & linguistic    & integer         & 0                           & 0                           \\
        8           & alphabet-g       &                  &                 &                  & \cmark             & \cmark                    & linguistic    & integer         & 0                           & 0                           \\
        9           & alphabet-h       &                  &                 &                  & \cmark             & \cmark                    & linguistic    & integer         & 0                           & 1                           \\
        10          & alphabet-i       &                  &                 &                  & \cmark             & \cmark                    & linguistic    & integer         & 2                           & 0                           \\
        11          & alphabet-j       & \cmark           & \cmark          &                  & \cmark             & \cmark                    & linguistic    & integer         & 0                           & 0                           \\
        12          & alphabet-k       &                  & \cmark          &                  & \cmark             & \cmark                    & linguistic    & integer         & 0                           & 2                           \\
        13          & alphabet-l       & \cmark           &                 &                  & \cmark             & \cmark                    & linguistic    & integer         & 0                           & 0                           \\
        14          & alphabet-m       & \cmark           & \cmark          &                  & \cmark             & \cmark                    & linguistic    & integer         & 0                           & 1                           \\
        15          & alphabet-n       & \cmark           & \cmark          &                  & \cmark             & \cmark                    & linguistic    & integer         & 0                           & 0                           \\
        16          & alphabet-o       &                  &                 & \cmark           & \cmark             & \cmark                    & linguistic    & integer         & 0                           & 0                           \\
        17          & alphabet-p       &                  &                 &                  & \cmark             & \cmark                    & linguistic    & integer         & 0                           & 2                           \\
        18          & alphabet-q       & \cmark           &                 &                  & \cmark             & \cmark                    & linguistic    & integer         & 0                           & 0                           \\
        19          & alphabet-r       & \cmark           &                 &                  & \cmark             & \cmark                    & linguistic    & integer         & 1                           & 0                           \\
        20          & alphabet-s       & \cmark           & \cmark          &                  & \cmark             & \cmark                    & linguistic    & integer         & 1                           & 0                           \\
        21          & alphabet-t       & \cmark           &                 &                  & \cmark             & \cmark                    & linguistic    & integer         & 1                           & 0                           \\
        22          & alphabet-u       &                  &                 &                  & \cmark             & \cmark                    & linguistic    & integer         & 1                           & 0                           \\
        23          & alphabet-v       &                  &                 &                  & \cmark             & \cmark                    & linguistic    & integer         & 0                           & 1                           \\
        24          & alphabet-w       & \cmark           &                 &                  & \cmark             & \cmark                    & linguistic    & integer         & 0                           & 2                           \\
        25          & alphabet-x       & \cmark           &                 &                  & \cmark             & \cmark                    & linguistic    & integer         & 0                           & 0                           \\
        26          & alphabet-y       & \cmark           & \cmark          &                  & \cmark             & \cmark                    & linguistic    & integer         & 1                           & 0                           \\
        27          & alphabet-z       & \cmark           & \cmark          &                  &                    & \cmark                    & linguistic    & integer         & 0                           & 0                           \\
        28          & consecutive-consonant-ratio& \cmark           & \cmark          & \cmark           &                    & \cmark                    & linguistic    & rational        & 0.16667                     & 1.00000                     \\
        29          & consecutive-digit-ratio&                  &                 & \cmark           &                    & \cmark                    & linguistic    & rational        & 0.00000                     & 0.00000                     \\
        30          & consonant-to-vowel-ratio& \cmark           &                 & \cmark           &                    & \cmark                    & linguistic    & rational        & 0.83333                     & 10.00000                    \\
        31          & consonants-character-ratio& \cmark           & \cmark          & \cmark           &                    & \cmark                    & linguistic    & rational        & 0.41667                     & 1.00000                     \\
        32          & consonants-max-streak-length& \cmark           & \cmark          &                  &                    & \cmark                    & linguistic    & integer         & 2                           & 10                          \\
        33          & contains-digits  &                  &                 & \cmark           &                    &                           & linguistic    & binary          & 0                           & 0                           \\
        34          & decimaldigits-character-ratio& \cmark           & \cmark          & \cmark           &                    & \cmark                    & linguistic    & rational        & 0.00000                     & 0.00000                     \\
        35          & decimaldigits-max-streak-length& \cmark           &                 & \cmark           &                    &                           & linguistic    & integer         & 0                           & 0                           \\
        36          & end-digit-edge-distance& \cmark           &                 & \cmark           & \cmark             &                           & linguistic    & integer         & -1                          & -1                          \\
        37          & first-character-pair& \cmark           & \cmark          &                  & \cmark             & \cmark                    & linguistic    & integer         & 10411                       & 14379                       \\
        38          & hexadecimaldigits-character-ratio& \cmark           & \cmark          & \cmark           &                    & \cmark                    & linguistic    & rational        & 0.33333                     & 0.10000                     \\
        39          & hexadecimaldigits-max-streak-length&                  & \cmark          & \cmark           &                    & \cmark                    & linguistic    & integer         & 2                           & 1                           \\
        40          & inverse-hamming-distance& \cmark           &                 &                  & \cmark             & \cmark                    & linguistic    & rational        & 1.00000                     & 1.00000                     \\
        41          & primedigits-character-ratio& \cmark           &                 &                  & \cmark             & \cmark                    & linguistic    & rational        & 0.08333                     & 0.30000                     \\
        42          & primedigits-max-streak-length& \cmark           & \cmark          &                  &                    & \cmark                    & linguistic    & integer         & 1                           & 2                           \\
        43          & repeated-characters-ratio&                  &                 & \cmark           & \cmark             &                           & linguistic    & rational        & 0.22222                     & 0.42857                     \\
        44          & start-digit-edge-distance& \cmark           & \cmark          & \cmark           & \cmark             & \cmark                    & linguistic    & integer         & -1                          & -1                          \\
        45          & subdomain-digit-sum& \cmark           & \cmark          &                  & \cmark             & \cmark                    & linguistic    & integer         & 238                         & 237                         \\
        46          & suffix-digit-sum & \cmark           & \cmark          & \cmark           & \cmark             & \cmark                    & linguistic    & integer         & 67                          & 80                          \\
        47          & suffix-standard-deviation& \cmark           & \cmark          & \cmark           & \cmark             & \cmark                    & linguistic    & rational        & 4.64280                     & 3.67423                     \\
        48          & syllable-count   & \cmark           &                 & \cmark           & \cmark             & \cmark                    & linguistic    & integer         & 4                           & 2                           \\
        49          & vowels-character-ratio& \cmark           &                 & \cmark           & \cmark             & \cmark                    & linguistic    & rational        & 0.50000                     & 0.00000                     \\
        50          & vowels-max-streak-length& \cmark           & \cmark          & \cmark           & \cmark             & \cmark                    & linguistic    & integer         & 3                           & 0                           \\
        51          & weighted-streaks & \cmark           &                 &                  &                    & \cmark                    & linguistic    & rational        & 0.65278                     & 20.36000                    \\
        52          & domain-name-length& \cmark           & \cmark          & \cmark           & \cmark             & \cmark                    & structural    & integer         & 16                          & 15                          \\
        53          & hex-exclusive-subdomains-ratio& \cmark           &                 & \cmark           & \cmark             & \cmark                    & structural    & rational        & 0.00000                     & 0.00000                     \\
        54          & second-level-length& \cmark           & \cmark          & \cmark           & \cmark             & \cmark                    & structural    & integer         & 12                          & 10                          \\
        55          & subdomains-length& \cmark           & \cmark          & \cmark           & \cmark             & \cmark                    & structural    & integer         & 12                          & 10                          \\
        56          & subdomains-mean-length& \cmark           & \cmark          & \cmark           & \cmark             & \cmark                    & structural    & rational        & 12.00000                    & 10.00000                    \\
        57          & suffix-length    & \cmark           & \cmark          & \cmark           & \cmark             & \cmark                    & structural    & integer         & 3                           & 4                           \\
        58          & 1-gram-alphabet-diversity& \cmark           &                 & \cmark           & \cmark             & \cmark                    & statistical   & rational        & 0.75000                     & 0.70000                     \\
        59          & 1-gram-alphabet-size& \cmark           &                 & \cmark           & \cmark             &                           & statistical   & integer         & 9                           & 7                           \\
        60          & 1-gram-arithmetic-mean& \cmark           &                 &                  &                    &                           & statistical   & rational        & 1.33333                     & 1.42857                     \\
        61          & 1-gram-harmonic-mean& \cmark           &                 & \cmark           &                    & \cmark                    & statistical   & rational        & 1.14894                     & 1.27273                     \\
        62          & 1-gram-kurtosis   &                  &                 &                  & \cmark             & \cmark                    & statistical   & rational        & 1.50000                     & -1.91667                    \\
        63          & 1-gram-max        &                  &                 & \cmark           & \cmark             & \cmark                    & statistical   & integer          & 3                           & 2                           \\
        64          & 1-gram-shannon-entropy& \cmark           &                 & \cmark           & \cmark             & \cmark                    & statistical   & rational        & 3.02206                     & 2.72193                     \\
        65          & 1-gram-skewness   &                  &                 &                  & \cmark             & \cmark                    & statistical   & rational        & 1.75000                     & 0.28868                     \\
        66          & 1-gram-standard-deviation& \cmark           &                 &                  &                    & \cmark                    & statistical   & rational        & 0.66667                     & 0.49487                     \\
        67          & 2-gram-alphabet-size& \cmark           &                 & \cmark           & \cmark             &                           & statistical   & integer         & 11                          & 9                           \\
        68          & 2-gram-shannon-entropy& \cmark           & \cmark          & \cmark           & \cmark             &                           & statistical   & rational        & 3.45943                     & 3.16993                     \\
        69          & 3-gram-alphabet-size& \cmark           &                 & \cmark           & \cmark             &                           & statistical   & integer         & 10                          & 8                           \\
        70          & 3-gram-shannon-entropy& \cmark           &                 & \cmark           & \cmark             &                           & statistical   & rational        & 3.32193                     & 3.00000                     \\
        71          & binary-matrix-rank-test&                  &                 & \cmark           & \cmark             & \cmark                    & statistical   & binary          & 1                           & 0                           \\
        72          & binary-matrix-rank-test-unicode&                  &                 & \cmark           & \cmark             & \cmark                    & statistical   & binary          & 0                           & 1                           \\
        73          & bits-entropy     & \cmark           &                 & \cmark           & \cmark             &                           & statistical   & rational        & 5.67243                     & 5.45943                     \\
        74          & longest-run-of-ones-test&                  &                 & \cmark           & \cmark             & \cmark                    & statistical   & binary          & 1                           & 1                           \\
        75          & longest-run-of-ones-test-unicode&                  &                 &                  & \cmark             & \cmark                    & statistical   & binary          & 1                           & 1                           \\
        76          & zlib-bits-compression-ratio& \cmark           &                 & \cmark           &                    &                           & statistical   & rational        & 2.28571                     & 2.42424                     \\
        \bottomrule
    \end{tabular}
}
\end{table*}

\end{document}